\documentstyle[amsfonts,12pt]{article}

\input{tcilatex}

\begin{document}

\title{Analytical technique for simplification of the encoder-decoder
circuit for a perfect five-qubit error correction}
\author{Jin-Yuan Hsieh$^{1}$, Che-Ming Li$^{2}$, and Der-San Chuu$^{2}$ \\
$^{1}$Department of Mechanical Engineering, Ming Hsin University \\
of Science and Technology, Hsinchu 30401,Taiwan.\\
$^{2}$Institute and Department of Electrophysics, National Chiao\\
Tung University, Hsinchu 30050, Taiwan.}
\maketitle

\begin{abstract}
Simpler encoding and decoding networks are necessary for more reliable
quantum error correcting codes (QECCs). The simplification of the
encoder-decoder circuit for a perfect five-qubit QECC can be derived
analytically if the QECC is converted from its equivalent one-way
entanglement purification protocol (1-EPP). In this work, the analytical
method to simplify the encoder-decoder circuit is introduced and a circuit
that is as simple as the existent simplest circuits is presented as an
example. The encoder-decoder circuit presented here involves nine single-
and two-qubit unitary operations, only six of which are controlled-NOT
(CNOT) gates.

\end{abstract}

\section{Introduction\protect\bigskip}

The unique feature of quantum correlation and quantum interference has
stimulated ingenious scenarios to exhibit the power of quantum information
processing \cite{qi}. Quantum states can be encoded into qubits through {\it %
quantum error-correcting codes}. With an introduction of redundancy, the
encoded data can tolerate little errors which are due to decoherence in some
individual qubits. Then, quantum error-correcting codes play a crucial role
in scalable quantum computation and communication to preserve the gain in
computational time and in security.

The five-qubit quantum error-correcting code (QECC) that protects a qubit of
information against general one-qubit errors is one of special interests for
quantum computations. It has been proven to be the best and smallest block
code \cite{kl}. It is also a perfect non-degenerate code because it
saturates the quantum Hamming bound \cite{em} and thus is capable of
correcting all one-qubit errors with minimum number of extra qubits.
Laflamme et al. \cite{l} and Bennett {\it et al}. \cite{b} independently
showed the first five-qubit QECCs. Recent developments of most QECCs are
attributed to stabilizer formalisms \cite{g,c}. In the work of Laflamme {\it %
et al}. \cite{l}, the five-qubit error correction is described to perform in
a rather simple procedure. The initial one-qubit information, as accompanied
with four extra qubits in the state $\left| 0\right\rangle $, is encoded by
a circuit representing a sequence of single-qubit Pauli operations and
two-qubit controlled Pauli operations. Then, after the interaction of
environment that causes generic one-qubit errors, the polluted five-qubit
state is decoded by running the same encoder circuit in a reverse order.
Eventually, the tensor product state of the four extra qubits is measured in
the computational basis ($\left| 0\right\rangle $ and $\left| 1\right\rangle 
$) to decide the corresponding final Pauli operation for recovering the
original state of the information carried qubit. By computer search,
Braunstein and Smolin \cite{bs} found a simplified encoder circuit which can
encode the one-qubit information in 24 laser pulses. For the stabilizer
code, however, the initial one-qubit information is encoded by the actions
of all the operators belonging to the group generated by the stabilizers.
The encoded five-qubit state is then allowed to be affected by generic
one-qubit errors followed by measurements of the stabilizer observables to
detect and correct the qubit on which the error has occurred. The fiv-qubit
stabilizer code has been experimentally implemented using nuclear magnetic
resonance by Knill {\it et al}. \cite{k}. The five-qubit QECC introduced by
Bennett {\it et al}. \cite{b} was derived from a restricted one-way
entanglement purification protocol (1-EPP) which purifies one good Bell
state from a noisy block of five Bell states. In fact, it can be shown that
the Bennett{\it \ et al}. protocol is equivalent to the error correction of
Laflamme et al. However, the QECC of Bennett {\it et al}. can be well
derived so that it requires a simpler network for both encoding and decoding
than the original one reported by Laflamme {\it et al}.. Bennett {\it et al}%
. suggested to use a Monte Carlo search program for deriving the QECC.

In realistic situations, to reduce the number of two-qubit gates necessary
in the encoder-decoder circuit is significantly important for reliable
five-qubit QECCs because two-qubit operations could be the more difficult
ones to be implemented in a physical apparatus \cite{bb}. This work thus is
motivated to derive five-qubit, single-error corrections which can be
performed by using the least number of two-qubit operations in their
encoder-decoder networks. The QECC presented as an example herein is derived
analytically from the restricted 1-EPP proposed by Bennett {\it et al}. \cite%
{b} and its encoder-decoder network contains only six controlled-NOT (CNOT)
gates and three single-qubit operations. The restricted 1-EPP therefore is
depicted first in the next section. In section 3, we describe the systematic
method for deriving 1-EPP in detail. A concrete example for the simplest
quantum gate array then will be given to show the capacity of the present
method. In section 4, we present the coding circuit which is converted
directly from the 1-EPP and compare its efficiency with those of several
existent encoder-decoder circuits. A conclusion is given in the final
section.

\section{The 5-EPR-pair single-error-correcting code\protect\bigskip}

Suppose there exists a finite block-size 1-EPP which distills one good pair
of spins in a specific Bell state from a block of five pairs, and no more
than one of the five pairs is subjected to noise. When this 1-EPP is
combined with a teleportation protocol, two parties, Alice and Bob, can
transmit quantum states reliably from one to the other. The combination of
the 1-EPP and teleportation protocol therefore is equivalent to a QECC. The
1-EPP considered herein is schematically depicted in Fig. 1. Suppose Alice
is the encoder, Bob the decoder, and the Bell state $\Phi ^{+}=(\left|
00\right\rangle +\left| 11\right\rangle )/\sqrt{2}$ is the good state to be
purified. Alice and Bob are supposed to be provided with five pairs of spins
in the state $\Phi ^{+}$ by a quantum source (QS). However, they actually
share five Bell states in which generic errors have or have not occurred on
at most one Bell state due to the presence of noise N$_{B}$ in the quantum
channel via which the pairs are transmitted. The noise models are assumed to
be one-sided \cite{b} and can cause the good Bell state $\Phi ^{+}$ to
become one of the incorrect Bell states 
\begin{equation}
\Phi ^{-}=\frac{1}{\sqrt{2}}(\left| 00\right\rangle -\left| 11\right\rangle
),\Psi ^{\pm }=\frac{1}{\sqrt{2}}(\left| 01\right\rangle \pm \left|
10\right\rangle ).
\end{equation}%
The good Bell state $\Phi ^{+}$ can become one of the erroneous Bell states
expressed in (1) if it is subjected to either a phase error ($\Phi
^{+}\rightarrow \Phi ^{-}$), an amplitude error ($\Phi ^{+}\rightarrow \Psi
^{+}$), or both ($\Phi ^{+}\rightarrow \Psi ^{-}$)\cite{kl,s}. When
performing the 1-EPP, Alice and Bob have a total of 16 error syndromes to
deal with. The collection of error syndromes includes the case that none of
the five pairs has been subjected to errors and the 15 cases in which one of
the five pairs has been subjected to one of the three types of error. The
strategy of Alice and Bob is to perform a sequence of unilateral and
bilateral unitary operations (as shown in Fig. 1, $U_{1}$ and $U_{2}$
performed by Alice and Bob, respectively) to transform the collection of the
16 error syndromes to another collection that can provide information about
the errors subjected by their particles. Suppose the state of the first pair
in the block is to be recovered. After performing the sequence of their
operations ($U_{1}$ and $U_{2}$ respectively), Alice and Bob, should then
perform local measurements on their respective halves of the second to fifth
pairs. Alice sends her result via classical channels to Bob who then
performs the Pauli operation $U_{3}$ to recover the original state of the
first pair conditionally on both Alice's and his results. The ultimate
requirement of these results of final measurement is that each and every of
them should be distinguishable from the others. In other words, there should
be 16 distinct measurements obtained from the aforementioned transformation
of the error syndrome. The main issue now is that the sequence of unilateral
and bilateral unitary operations performed by the two parties to transform
the error syndrome should be well designed so the requirement just mentioned
can be fulfilled.

To arrange the sequence of operations, basic concepts of linear algebra are
used. The four Bell states $\Phi ^{\pm }$ and $\Psi ^{\pm }$ are first
labeled by two classical bits, namely, 
\begin{equation}
\Phi ^{+}=00,\Phi ^{-}=10,\Psi ^{+}=01,\Psi ^{-}=11.
\end{equation}%
The right, low-order or amplitude bit identifies the $\Phi /\Psi $ property
of the Bell state, while the left, high-order or phase bit identifies the $%
+/-$ property. Note that the combined result of the local measurements
obtained by Alice and Bob on a Bell state is revealed by the Bell state's
low or amplitude bit. In the representation of the high-low bits, each error
syndrome thus is expressed as a ten-bit codeword, e.g., the error syndrome $%
\Phi ^{+}\Psi ^{-}\Phi ^{+}\Phi ^{+}\Phi ^{+}$ is written as $00$ $11$ $00$ $%
00$ $00$. Codewords of the error syndrome, denoted by $e_{r}^{(i)},$ $%
i=0,1,...,15,$ are listed in Table 1. The effect of the sequence of
unilateral and bilateral unitary operations performed by Alice and Bob is to
map the codewords $e_{r}^{(i)}$ onto another collection of ten-bit codewords 
$w^{(i)}.$ If both the codewords $e_{r}^{(i)}$ and $w^{(i)}$ are written as
column vectors in the ten-dimensional Boolean-valued ($\in \{0,1\}$) space,
then the mapping $e_{r}^{(i)}\rightarrow w^{(i)}$ can be simply expressed by
a matrix equation 
\begin{equation}
w^{(i)}={\bf M}e_{r}^{(i)},
\end{equation}%
provided that the mapping is confined to $w^{(0)}=e_{r}^{(0)}(=00$ $00$ $00$ 
$00$ $00).$ The four error syndromes, $e_{r}^{(3k)}$, $e_{r}^{(3k-1)}$, $%
e_{r}^{(3k-2)}$, and $e_{r}^{(0)}$, corresponding to a common erroneous
pair, form a group and are characterized by 
\begin{equation}
e_{r}^{(3k-2)}\oplus e_{r}^{(3k-1)}=e_{r}^{(3k)},k=1,2,...,5,
\end{equation}%
where $k$ enumerates the erroneous pair and $\oplus $ is the addition modulo
2. Accordingly, the 16 codewords $w^{(i)}$ should be subdivided into five
corresponding groups, each of which has $w^{(3k)}$, $w^{(3k-1)}$, $%
w^{(3k-2)} $, and $w^{(0)}$, and holds the relation 
\begin{equation}
w^{(3k-2)}\oplus w^{(3k-1)}=w^{(3k)},k=1,2,...,5.
\end{equation}%
Therefore the matrix ${\bf M}$ can be simply expressed by a $10\times 10$
matrix, such as 
\begin{equation}
{\bf M}=\left[
w^{(1)}w^{(2)}w^{(4)}w^{(5)}w^{(7)}w^{(8)}w^{(10)}w^{(11)}w^{(13)}w^{(14)}%
\right] ,
\end{equation}%
in accordance with the arrangement of error syndromes listed in Table 1. The
first two rows of ${\bf M}$ represent the states of the pair to be
recovered, and the 4$^{th}$, 6$^{th}$, 8$^{th}$,and 10$^{th}$ rows represent
the low bits of the second to fifth Bell states and thus construct the
four-bit codewords for the measurement results $v^{(i)}.$ The measurement
result $v^{(i)}$\ of course is also characterized by 
\begin{equation}
v^{(3k-2)}\oplus v^{(3k-1)}=v^{(3k)},k=1,2,...,5,
\end{equation}%
in accordance with relations (4) and (5). In the language of linear algebra,
the action of the sequence of unilateral and bilateral unitary operations
that accounts for the mapping $e_{r}^{(i)}\rightarrow w^{(i)}$ is to perform
a sequence of elementary row operations on the $10\times 10$ identity matrix 
${\bf 1}$ to reduce it to the matrix ${\bf M.}$ In this spirit, Bennett {\it %
et al. }\cite{b} have undertaken a Monte Carlo numerical search program to
find out suitable solutions for matrix ${\bf M}$ and their corresponding
encoder-decoder networks. Basically, the approach implemented by Bennett 
{\it et al.} is a tedious numerical method of trial and error performing the
transformation ${\bf 1\rightarrow M}$ subjected to a ''forward'' sequence of
local operations. In this work, we will present an analytical method for
creating ${\bf M}$ implemented in the present QECC. The present method will
be described in detail in the next section.

\section{The present method\protect\bigskip}

\subsection{Theory}

The unilateral and bilateral unitary operations performed in the 1-EPP in
fact are their own inverse transformations, so if the sequence of operations
is run in the reverse order, then the inverse transformations ${\bf %
M\rightarrow 1}$ is accomplished. In the spirit of inverse transformation,
it thus allows us to derive all appropriate versions of ${\bf M}$ and the
corresponding encoder-decoder networks by following an analytical way. More
importantly, for a derived ${\bf M}$, rearranging the sequence of row
operations on the same inverse transformation ${\bf M\rightarrow 1}$ will
help in constructing its simplest encoder-decoder network.

An elementary row operation corresponds to a basic unilateral or bilateral
unitary operation. In the present protocol, Alice and Bob are confined to
perform only three basic unitary operations because these operations are
necessary and sufficient for the elementary row operations needed to achieve
the mapping ${\bf M}\rightarrow {\bf 1}$, and vice versa. These basic
operations are: (1) a bilateral CNOT (BXOR), which performs the bit change $%
(x_{S}$, $y_{S})(x_{T},y_{T})\rightarrow (x_{S}\oplus x_{T}$, $%
y_{S})(x_{T},y_{S}\oplus y_{T})$, where the subscripts $S$ and $T$ denote
the source and target pairs, respectively; (2) a bilateral $\pi /2-$rotation 
$B_{y}$, which performs $(x,$ $y)\rightarrow (y,x)$; \ and (3) a composite
operation $\sigma _{x}B_{x}$, which performs $(x,$ $y)\rightarrow (x,x\oplus
y)$. The unitary Pauli operation $\sigma _{x}$ performs a $\pi $-rotation of
Alice or Bob's spin about the $x-$axis, while the bilateral operation $B_{x}$
$(B_{y})$ performs a $\pi /2-$rotation of both Alice and Bob's spins about
the $x$ $(y)-$axis. The unilateral operations are defined as those operators
performed by Alice or Bob but not both. The bilateral operations are
represented by a tensor product of one part of Bob and the same part of
Alice. Note that the bilateral CNOT is performed such that the source qubits
of Alice and Bob belong to a common pair, and the target qubits belong to
another common pair.

The information obtained through local measurements and one-way
communications can only deduce the low bit of a Bell pair, and the original
state of the first Bell pair can only be recovered by the low-bit
information. Then, for a successful 1-EPP, or its equivalent QECC, each and
every measurement result $v^{(i)}$ is required to be distinguishable from
the others, so the collection of $v^{(i)}$ in fact should contain all
elements in the 4-dimensional Boolean-valued space. To perform the
aforementioned inverse transformation ${\bf M\rightarrow 1,}$ the codewords
of measurement result are first arranged according to relations (7) and the
matrix ${\bf M}$ can be assumed as 
\begin{equation}
{\bf M}=\left[ 
\begin{array}{cccccccccc}
a_{1} & a_{2} & a_{3} & a_{4} & a_{5} & a_{6} & a_{7} & a_{8} & a_{9} & 
a_{10} \\ 
b_{1} & b_{2} & b_{3} & b_{4} & b_{5} & b_{6} & b_{7} & b_{8} & b_{9} & 
b_{10} \\ 
c_{1} & c_{2} & c_{3} & c_{4} & c_{5} & c_{6} & c_{7} & c_{8} & c_{9} & 
c_{10} \\ 
0 & 0 & 1 & 0 & 0 & 1 & 0 & 1 & 1 & 0 \\ 
d_{1} & d_{2} & d_{3} & d_{4} & d_{5} & d_{6} & d_{7} & d_{8} & d_{9} & 
d_{10} \\ 
0 & 1 & 0 & 1 & 1 & 0 & 0 & 0 & 1 & 0 \\ 
e_{1} & e_{2} & e_{3} & e_{4} & e_{5} & e_{6} & e_{7} & e_{8} & e_{9} & 
e_{10} \\ 
1 & 0 & 0 & 1 & 0 & 1 & 1 & 0 & 0 & 0 \\ 
f_{1} & f_{2} & f_{3} & f_{4} & f_{5} & f_{6} & f_{7} & f_{8} & f_{9} & 
f_{10} \\ 
1 & 1 & 1 & 1 & 0 & 1 & 0 & 0 & 0 & 1%
\end{array}%
\right] .
\end{equation}%
It should be noted that the arrangement of the results of measurements shown
in the above matrix is only one\ of the possible choices. By performing a
sequence of row operations corresponding to the basic unitary operations,
the assumed matrix ${\bf M}$ (8) actually is allowed to be reduced to one of
all the alternatives akin to the identity matrix ${\bf 1}$, and a suitable
encoder-decoder network is constructed accordingly. The alternatives akin to
the identity ${\bf 1}$ are those obtained by 1$-$ permuting column vectors
within one of the five sets of two column vectors ($x^{(3k-2)}$ and $%
x^{(3k-1)}$, $k=1,2,...,5$), or 2$-$ adding one column to the other within
each of the groups, or 3$-$ performing both actions. For example, an
alternative could be 
\begin{equation}
{\bf 1}_{akin}=\left[ 
\begin{array}{cccccccccc}
1 & 1 & 0 & 0 & 0 & 0 & 0 & 0 & 0 & 0 \\ 
0 & 1 & 0 & 0 & 0 & 0 & 0 & 0 & 0 & 0 \\ 
0 & 0 & 1 & 1 & 0 & 0 & 0 & 0 & 0 & 0 \\ 
0 & 0 & 0 & 1 & 0 & 0 & 0 & 0 & 0 & 0 \\ 
0 & 0 & 0 & 0 & 0 & 1 & 0 & 0 & 0 & 0 \\ 
0 & 0 & 0 & 0 & 1 & 0 & 0 & 0 & 0 & 0 \\ 
0 & 0 & 0 & 0 & 0 & 0 & 0 & 1 & 0 & 0 \\ 
0 & 0 & 0 & 0 & 0 & 0 & 1 & 0 & 0 & 0 \\ 
0 & 0 & 0 & 0 & 0 & 0 & 0 & 0 & 0 & 1 \\ 
0 & 0 & 0 & 0 & 0 & 0 & 0 & 0 & 1 & 0%
\end{array}%
\right] .
\end{equation}%
When the derivation of ${\bf M}$ is done, the alternative akin to ${\bf 1}$
is then converted back to the identity ${\bf 1}$ by well rearranging its
columns and the derived ${\bf M}$ is adjusted via the same column changes,
in order to conform equation (3). The procedure of reducing the matrix ${\bf %
M}$ to the alternative akin to the identity ${\bf 1}$ is similar to the
Gauss-Jordan elimination method for solving systems of linear equations.
During the procedure of row operations, all the unknowns appearing in the
assumed matrix ${\bf M}$ (8) are given or solved according to the structure
of the alternative akin to ${\bf 1.}$ Details of the derivation can be found
in Ref. \cite{hl}.

\subsection{A systematic scenario example}

There are so many solutions for the assumed ${\bf M}$ which are all suitable
for the 1-EPP, however, only one of them has been adjusted and presented as: 
\begin{equation}
{\bf M_{1}}=\left[ 
\begin{array}{cccccccccc}
a_{1} & a_{2} & a_{3} & a_{4} & a_{5} & a_{6} & a_{7} & a_{8} & a_{9} & 
a_{10} \\ 
b_{1} & b_{2} & b_{3} & b_{4} & b_{5} & b_{6} & b_{7} & b_{8} & b_{9} & 
b_{10} \\ 
c_{1} & c_{2} & c_{3} & c_{4} & c_{5} & c_{6} & c_{7} & c_{8} & c_{9} & 
c_{10} \\ 
0 & 0 & 1 & 1 & 1 & 0 & 1 & 0 & 0 & 1 \\ 
d_{1} & d_{2} & d_{3} & d_{4} & d_{5} & d_{6} & d_{7} & d_{8} & d_{9} & 
d_{10} \\ 
0 & 1 & 0 & 1 & 0 & 1 & 0 & 0 & 0 & 1 \\ 
e_{1} & e_{2} & e_{3} & e_{4} & e_{5} & e_{6} & e_{7} & e_{8} & e_{9} & 
e_{10} \\ 
1 & 1 & 0 & 1 & 1 & 0 & 0 & 1 & 0 & 0 \\ 
f_{1} & f_{2} & f_{3} & f_{4} & f_{5} & f_{6} & f_{7} & f_{8} & f_{9} & 
f_{10} \\ 
1 & 0 & 1 & 0 & 1 & 0 & 0 & 0 & 1 & 0%
\end{array}%
\right] .
\end{equation}%
Let us show the systematic scenario for accomplishing the transformation $%
{\bf M_{1}}{\bf \rightarrow 1}$ by one of the simplest networks. The matrix $%
{\bf M_{1}}$ can be rephrased as 
\[
{\bf M_{1}}=\left[ 
\begin{array}{cccc}
m_{11} & m_{12} & \cdots & m_{15} \\ 
m_{21} & m_{22} & \cdots & m_{25} \\ 
\vdots & \vdots &  & \vdots \\ 
m_{51} & m_{52} & \cdots & m_{55}%
\end{array}%
\right] , 
\]%
where the matrix elements $m_{\alpha \beta }$ denote the $2\times 2$
matrices: 
\begin{equation}
m_{11}=\left[ 
\begin{array}{cc}
a_{1} & a_{2} \\ 
b_{1} & b_{2}%
\end{array}%
\right] ,m_{21}=\left[ 
\begin{array}{cc}
c_{1} & c_{2} \\ 
0 & 0%
\end{array}%
\right] ,...,
\end{equation}%
and so forth. The next step of our method is a procedure of elementary row
operations on the matrix ${\bf M_{1}}$ (10) subjected to a suitable sequence
of the basic operations. When the assumed matrix ${\bf M_{1}}$ is
transformed into the identity matrix ${\bf 1}$ under the series of row
operations, the unknowns $a_{r}$, $b_{r}$, ..., $f_{r}$ will be solved
stepwise in accordance with the structure of ${\bf 1}$. It is easy to show
that a sequence of row operations can do the transformation on two Bell
states $\alpha $ and $\beta $ in a group enumerated by $\gamma $, namely, 
\begin{equation}
\left[ 
\begin{array}{c}
m_{\alpha \gamma } \\ 
m_{\beta \gamma }%
\end{array}%
\right] \rightarrow \left[ 
\begin{array}{c}
{\bf I} \\ 
0%
\end{array}%
\right] ,
\end{equation}%
provided that $\det (m_{\alpha \gamma })=1$ and $\det (m_{\beta \gamma })=0$
. Here ${\bf I}$ denotes the $2\times 2$ identity matrix. For example, the
consecutive transformation 
\[
\left[ 
\begin{array}{c}
m_{\alpha \gamma } \\ 
m_{\beta \gamma }%
\end{array}%
\right] =\left[ 
\begin{array}{c}
\begin{array}{cc}
1 & 0 \\ 
1 & 1%
\end{array}
\\ 
\begin{array}{cc}
0 & 1 \\ 
0 & 0%
\end{array}%
\end{array}%
\right] \rightarrow \left[ 
\begin{array}{c}
\begin{array}{cc}
1 & 0 \\ 
1 & 1%
\end{array}
\\ 
\begin{array}{cc}
0 & 0 \\ 
0 & 1%
\end{array}%
\end{array}%
\right] \rightarrow \left[ 
\begin{array}{c}
\begin{array}{cc}
1 & 0 \\ 
0 & 1%
\end{array}
\\ 
\begin{array}{cc}
0 & 0 \\ 
0 & 1%
\end{array}%
\end{array}%
\right] \rightarrow \left[ 
\begin{array}{c}
\begin{array}{cc}
1 & 0 \\ 
0 & 1%
\end{array}
\\ 
\begin{array}{cc}
0 & 0 \\ 
0 & 0%
\end{array}%
\end{array}%
\right] 
\]%
can be accomplished if the operation $B_{y}$ is first performed on Bell
state $\beta $, then a $\sigma _{x}B_{x}$ is performed on Bell state $\alpha 
$ followed by a BXOR performed on both states, as Bell state $\alpha $ being
the source and Bell state $\beta $ being the target. It can be found in what
follows that the unknowns assumed in the matrix ${\bf M_{1}}$ either will be
given based on the requirement for the transformation described in (13), or
will be determined according to the unique structure of the identity matrix $%
{\bf 1}$.

In the first stage of row operations, we are confined to performing a
transformation of the matrix ${\bf M_{1}}$ (11) such that $m_{44}\rightarrow 
{\bf I}$ and $m_{4k},$ $m_{k4}\rightarrow 0$, for $k=1,$ $2,$ $3,$ and $5$,
according to the structure of ${\bf 1}$. Let $\det (m_{44})=1$ and $\det
(m_{14})=...=\det (m_{54})=0,$ which imply 
\begin{equation}
a_{7}b_{8}\oplus
a_{8}b_{7}=0,\,c_{8}=0,\,e_{7}=1;\,c_{7},\,d_{7},\,d_{8},\,e_{8},\,f_{7},%
\,f_{8}\in \{0,1\}.
\end{equation}%
Clearly, there are totally 640 solutions for the unknowns appearing in (10)
to be considered in this stage. (10 for the condition $a_{7}b_{8}\oplus
a_{8}b_{7}=0$, 2 for each of the 6 arbitrary Boolean valued unknowns, and
thus totally $10\times 2^{6}=640$ solutions) To illustrate the simplest way
of creating Boolean functions, however, only one among these 640 cases is
considered. Let us consider the case in which 
\begin{equation}
a_{7}=1,\,b_{7}=a_{8}=b_{8}=c_{7}=d_{7}=d_{8}=e_{8}=f_{7}=f_{8}=0.
\end{equation}%
Then, by performing the operations shown in Fig. 2(a), we have the
transformation ${\bf M_{1}}\rightarrow {\bf M_{1}^{\prime }}$, 
\begin{eqnarray}
{\bf M_{1}^{\prime }} &=&\left[ 
\begin{array}{cccccccccc}
a_{1} & a_{2} & a_{3} & a_{4} & a_{5} & a_{6} & 0 & 0 & a_{9} & a_{10} \\ 
1 & 1 & 0 & 0 & 1 & 0 & 0 & 0 & 0 & 1 \\ 
0 & 0 & 1 & 0 & 1 & 0 & 0 & 0 & 0 & 0 \\ 
0 & 0 & 0 & 1 & 0 & 0 & 0 & 0 & 0 & 1 \\ 
d_{1} & d_{2} & d_{3} & d_{4} & d_{5} & d_{6} & 0 & 0 & d_{9} & d_{10} \\ 
0 & 1 & 0 & 1 & 0 & 1 & 0 & 0 & 0 & 1 \\ 
0 & 0 & 0 & 0 & 0 & 0 & 1 & 0 & 0 & 0 \\ 
0 & 0 & 0 & 0 & 0 & 0 & 0 & 1 & 0 & 0 \\ 
f_{1} & f_{2} & f_{3} & f_{4} & f_{5} & f_{6} & 0 & 0 & f_{9} & f_{10} \\ 
1 & 0 & 1 & 0 & 1 & 0 & 0 & 0 & 1 & 0%
\end{array}%
\right]  \nonumber \\
&=&\left[ 
\begin{array}{ccccc}
m_{11}^{\prime } & m_{12}^{\prime } & m_{13}^{\prime } & 0 & m_{14}^{\prime }
\\ 
m_{21}^{\prime } & m_{22}^{\prime } & m_{23}^{\prime } & 0 & m_{25}^{\prime }
\\ 
m_{31}^{\prime } & m_{32}^{\prime } & m_{33}^{\prime } & 0 & m_{35}^{\prime }
\\ 
0 & 0 & 0 & {\bf I} & 0 \\ 
m_{51}^{\prime } & m_{52}^{\prime } & m_{53}^{\prime } & 0 & m_{55}^{\prime }%
\end{array}%
\right] ,
\end{eqnarray}%
in which we have chosen the following setting for the unknowns: 
\begin{eqnarray}
&&b_{1}=1,b_{2}=1,b_{3}=0,b_{4}=0,b_{5}=1,b_{6}=b_{9}=0,b_{10}=1,  \nonumber
\\
&&c_{1}=0,c_{2}=c_{3}=0,c_{4}=1,c_{5}=c_{6}=c_{9}=0,c_{10}=1,  \nonumber \\
&&and\quad e_{1}=e_{2}=e_{3}=e_{4}=e_{5}=e_{6}=e_{9}=e_{10}=0.
\end{eqnarray}

Let us proceed to apply the second series of operations, as depicted in the
Fig. 2(b), to perform the transformations $m_{22}^{\prime }\rightarrow {\bf I%
}$ and $m_{2k}^{\prime },$ $m_{k2}^{\prime }\rightarrow 0$, for $k=1,$ $3,$
and $5$. As a result, we have 
\begin{eqnarray}
&&d_{1}=f_{1}=d_{2}=f_{2}=0,\,d_{3}=d_{4}=f_{3}=f_{4}=0,d_{5}=1,%
\,d_{6}=0=f_{5}=f_{6}=0,\,  \nonumber \\
&&d_{9}=f_{9}=d_{10}=0,\,f_{10}=1,\,a_{3}=a_{4}=0.
\end{eqnarray}%
Note that according to the requirements det($m_{2k}^{\prime }$)=$0$ and det($%
m_{k2}^{\prime }$)=$0$, $a_{3}=a_{4}=0$ is only one of the suitable choices
and $d_{3}=d_{4}=0$ is the only choice. Therefore, the ${\bf M_{1}^{\prime }}
$ is transformed into ${\bf M_{1}^{\prime \prime }}$: 
\begin{eqnarray}
{\bf M_{1}^{\prime \prime }} &=&\left[ 
\begin{array}{cccccccccc}
a_{1} & a_{2} & 0 & 0 & a_{5} & a_{6} & 0 & 0 & a_{9} & a_{10} \\ 
1 & 1 & 0 & 0 & 1 & 0 & 0 & 0 & 0 & 1 \\ 
0 & 0 & 1 & 0 & 0 & 0 & 0 & 0 & 0 & 0 \\ 
0 & 0 & 0 & 1 & 0 & 0 & 0 & 0 & 0 & 0 \\ 
0 & 0 & 0 & 0 & 1 & 0 & 0 & 0 & 0 & 0 \\ 
0 & 1 & 0 & 0 & 0 & 1 & 0 & 0 & 0 & 0 \\ 
0 & 0 & 0 & 0 & 0 & 0 & 1 & 0 & 0 & 0 \\ 
0 & 0 & 0 & 0 & 0 & 0 & 0 & 1 & 0 & 0 \\ 
1 & 0 & 0 & 0 & 1 & 0 & 0 & 0 & 1 & 0 \\ 
0 & 0 & 0 & 0 & 0 & 0 & 0 & 0 & 0 & 1%
\end{array}%
\right]  \nonumber \\
&=&\left[ 
\begin{array}{ccccc}
m_{11}^{\prime \prime } & 0 & m_{13}^{\prime \prime } & 0 & m_{14}^{\prime
\prime } \\ 
0 & {\bf I} & 0 & 0 & 0 \\ 
m_{31}^{\prime \prime } & 0 & m_{33}^{\prime \prime } & 0 & m_{35}^{\prime
\prime } \\ 
0 & 0 & 0 & {\bf I} & 0 \\ 
m_{51}^{\prime \prime } & 0 & m_{53}^{\prime \prime } & 0 & m_{55}^{\prime
\prime }%
\end{array}%
\right] .
\end{eqnarray}

Finally, if the matrix ${\bf M_{1}^{\prime \prime }}$ is transformed through
additional two BXOR and one $\sigma _{x}B_{x}$ operations, as shown in Fig.
2(c), it results to the identity matrix ${\bf 1}$. In this stage, we have
set the rest of the unknowns to be one of the alternatives: $%
a_{1}=1,a_{2}=0,a_{5}=1,a_{6}=0,a_{9}=0,$ and $a_{10}=0$. The whole sequence
of basic operations, as shown in Fig. 3, is obtained by combining the three
sub-sequences as shown in Figs. 2(a)-(c). It will transform the matrix ${\bf %
M_{1}}$ into the identity matrix ${\bf 1}$. This network is the simplest one
since it involves only six BXORs, and the corresponding matrix reads 
\begin{equation}
{\bf M_{1}}=\left[ 
\begin{array}{cccccccccc}
1 & 0 & 0 & 0 & 1 & 0 & 1 & 0 & 0 & 0 \\ 
1 & 1 & 0 & 0 & 1 & 0 & 0 & 0 & 0 & 1 \\ 
0 & 0 & 0 & 1 & 0 & 0 & 0 & 0 & 0 & 1 \\ 
0 & 0 & 1 & 1 & 1 & 0 & 1 & 0 & 0 & 1 \\ 
0 & 0 & 0 & 0 & 1 & 0 & 0 & 0 & 0 & 0 \\ 
0 & 1 & 0 & 1 & 0 & 1 & 0 & 0 & 0 & 1 \\ 
0 & 0 & 0 & 0 & 0 & 0 & 1 & 0 & 0 & 0 \\ 
1 & 1 & 0 & 1 & 1 & 0 & 0 & 1 & 0 & 0 \\ 
0 & 0 & 0 & 0 & 0 & 0 & 0 & 0 & 0 & 1 \\ 
1 & 0 & 1 & 0 & 1 & 0 & 0 & 0 & 1 & 0%
\end{array}%
\right] .
\end{equation}%
Performed by this network, the correspondence between the error syndromes $%
e_{r}^{(i)}$ and the combined measurement results $v_{r}^{(i)}$ is also
listed in Table 1. Referring to Table 1, or the matrix ${\bf M}_{1}$, when
Bob obtains the measurement result $v^{(2)}(=0110),$ for example, he knows
the pair to be purified is in the state $\Psi ^{+}(=01)$ and thus simply
performs the Pauli operation $U_{3}^{(2)}=\sigma _{x}$ to recover it to the
good state $\Phi ^{+}$.

\section{The encoder-decoder circuit for a perfect five-qubit error
correction\protect\bigskip}

The 1-EPP depicted above can be directly converted to a five-qubit QECC
whose encoder-decoder circuit has the same configuration as the one shown in
Fig. 4. However, in the language of QECC, the classical high-low or
phase-amplitude bits used to code the Bell state in the 1-EPP are now used
to code operators belonging to the Pauli group, namely, ${\bf I}=00, \sigma
_{x}=01, \sigma _{z}=10, \sigma _{y}=11$. When acting on a single qubit, the
Pauli operator produces either no error (by ${\bf I}$), a bit flip error (by 
$\sigma _{x}$), a phase flip error (by $\sigma _{z}$), or a bit-phase flip
error (by $\sigma _{y}$). Therefore, such a code is convenient because the
codewords $e_{r}^{(i)}$ are now replaced by $E_{r}^{(i)}$, which represent
the 16 error syndromes described by five-Pauli-operartor tensor products.
Furthermore, the transformation described by the matrix equation (3) is now
replaced by the similarity transformation of operators described as: $%
W^{(i)}=UE_{r}^{(i)}U^{+}$, where $U$ $(U^{+})$ represents the sequence of
the basic operations performed in the decoder (encoder) circuit. Clearly,
both the encoder and decoder circuits have exactly the same quantum gate
arrangement but they should be run in opposite orders. In order to perform
the transformation mentioned above, this time the single-qubit Hadamard
transformation: $H=H^{+}=(\sigma _{x}+\sigma _{z})/\sqrt{2} $, is used to
perform the bit change $H(x,y)H^{+}\rightarrow (y,x),$ the single-qubit
transformation: $Q=Q^{+}=(\sigma _{y}+\sigma _{z})/\sqrt{2} $, is used to
perform $Q(x,y)Q^{+}\rightarrow (x,x\oplus y)$, and the two-qubit CNOT gate
is used to perform (CNOT)$(x_{S}$, $y_{S})(x_{T},y_{T})$(CNOT)$%
^{+}\rightarrow (x_{S}\oplus x_{T}$, $y_{S})(x_{T},y_{S}\oplus y_{T})$,
respectively. That is, in the five-qubit QECC to be presented the basic
single- and two-qubit operations needed to be implemented are $H,$ $Q$, and
CNOT.

For the present five-qubit QECC, the correspondence between the codewords $%
W^{(i)}$ and $E_{r}^{(i)}$ is exactly the same as that between the derived
matrix ${\bf M}_{1}$ given in (9) and the identity ${\bf 1}$. The QECC is
performed as follows. If a state $\left| \phi\right\rangle =\alpha \left|
0\right\rangle +\beta \left| 1\right\rangle $ is to be protected in a
quantum computation, it is first accompanied with four extra qubits in the
state $\left| 0\right\rangle $. Then the five-qubit state $\left|
\phi\right\rangle \left| 0\right\rangle \left| 0\right\rangle \left|
0\right\rangle \left| 0\right\rangle $ is encoded by the performance of $%
U^{+}$. After the encoded state is subjected to $E_{r}^{(i)},$ the erroneous
state then is decoded by the implementation of $U$. The resulting state
turns out to be 
\begin{eqnarray}
\left| \phi _{r}^{(i)}\right\rangle &=&UE_{r}^{(i)}U^{+}(\left|
\phi\right\rangle \left| 0\right\rangle \left| 0\right\rangle \left|
0\right\rangle \left| 0\right\rangle )  \nonumber \\
&=&W^{(i)}(\left| \phi\right\rangle \left| 0\right\rangle \left|
0\right\rangle \left| 0\right\rangle \left| 0\right\rangle )  \nonumber \\
&=&(U_{3}^{(i)}\left| \phi\right\rangle )\left| a^{\prime }\right\rangle
\left| b^{\prime }\right\rangle \left| c^{\prime }\right\rangle \left|
d^{\prime }\right\rangle ,
\end{eqnarray}
where $U_{3}^{(i)}$\ is the single-qubit Pauli operation acting on the first
qubit and is dependent on the measurement result on the four extra qubits.
When the extra qubits are measured in the computational basis, the
measurement result $v^{(i)}=a^{\prime }b^{\prime }c^{\prime }d^{\prime }$ is
obtained. Eventually, the corresponding Pauli operation $U_{3}^{(i)}$ is
performed on the remaining qubit, \ which is in the state $U_{3}^{(i)}\left|
\phi\right\rangle ,$\ to recover the initial state $\left| \phi\right\rangle
.$ The procedure of performing the five-qubit QECC is quite simple, same as
the one reported by Laflamme et al. \cite{l}, and is displayed schematically
in Fig. 4. The present QECC is equivalent to the aforementioned 1-EPP, which
adopts the network shown in Fig. 4, so Table 1 is also useful to it. As a
result, when referring to Table 1 again, if the measurement result $%
v^{(2)}=0110$ is read, then $U_{3}^{(2)}=\sigma _{x}$ is performed to
recover the initial state $\left| \phi\right\rangle =\alpha \left|
0\right\rangle +\beta \left| 1\right\rangle .$ The encoder-decoder circuit
required to perform the present QECC, as shown in Figs. 4(a) and (b), is
rather simple; it contains nine operations, in which only six CNOTs are
required. As a matter of fact, this circuit is one of the simplest ones
derived so far. The other best known circuit is the one presented by
Braunstein and Smolin\cite{bs} and its corresponding matrix is 
\begin{equation}
{\bf M}_{BS}=\left[ 
\begin{array}{cccccccccc}
1 & 0 & 1 & 0 & 1 & 0 & 0 & 0 & 0 & 0 \\ 
0 & 1 & 0 & 1 & 0 & 1 & 0 & 0 & 0 & 0 \\ 
0 & 0 & 1 & 0 & 1 & 0 & 0 & 0 & 0 & 0 \\ 
0 & 1 & 1 & 1 & 1 & 0 & 0 & 0 & 0 & 1 \\ 
1 & 0 & 1 & 0 & 0 & 0 & 1 & 0 & 0 & 0 \\ 
1 & 0 & 1 & 1 & 0 & 1 & 1 & 0 & 0 & 0 \\ 
0 & 0 & 0 & 0 & 0 & 0 & 1 & 0 & 0 & 0 \\ 
0 & 0 & 0 & 1 & 0 & 1 & 0 & 1 & 0 & 1 \\ 
0 & 0 & 1 & 0 & 1 & 0 & 1 & 0 & 1 & 0 \\ 
0 & 0 & 1 & 0 & 1 & 0 & 1 & 0 & 1 & 1%
\end{array}
\right] .
\end{equation}

The efficiency of a coding scheme can be charactered by the shortness of the
encoder-decoder circuit. The shortness criterion is based on the fewest
total operations or the fewest CNOT operations \cite{b}. The total
operations include one-qubit rotations and CNOTs. It is equivalent to
determine the minimum experimental efforts for implementing the shortest
coding circuit on a quantum computer. The number of laser pulses required to
perform a encoder-decoder circuit is a reasonable measure of the efficiency
for ion-trap computers \cite{bs,bc}. A qubit is coded through the ground
state and the long-lived excited state of an ion in an ion-trap quantum
computer \cite{cz}. The physical states are driven by laser beams to
implement the quantum logic gates further. To count the number of laser
pulses, the encoder circuit from Fig. 4(a) is rewritten in terms of the gate
primitives of an ion-trap quantum computer and shown in Fig. 4(c). It is
interesting to observe that two pairs of CNOTs (the 2$^{nd}$ and 3$^{rd}$
and the 4$^{th}$ and 5$^{th}$ ones) in the present circuit can be combined
as two three qubit gates and can be implemented as single element. Besides,
the functions of operators $U$ and $V$ implemented on an ion-trap quantum
computer are equivalent to the ones of operators $H$ and $Q$ respectively.
Since each single-qubit operation requires one laser pulse, the two-qubit
gate needs three pulses, and the three-qubit gate requires four laser
pulses, the present circuit also requires only 24 laser pulses if it is
implemented on an ion-trap quantum computer, same as the Braunstein and
Smolin circuit. The numbers of total operations, CNOTs, and laser pulses for
the circuits presented by Bennett {\it et al.} \cite{b} and Braunstein and
Smolin \cite{bs} have also been summarized in Table 2.

\section{Conclusion\protect\bigskip}

This work has presented a rather simple encoder-decoder circuit to perform
the five-qubit, single-error correction protocol. The QECC derived herein is
converted directly from the restricted 1-EPP depicted above, so a major part
of this work is dedicated to the depiction of the 1-EPP. The present
encoder-decoder circuit is the simplest one corresponding to the derived
matrix ${\bf M_{1}}$ given in (20), which is derived via an analytical
approach \cite{hl}. This analytical approach, as shown, can help in deriving
not only the suitable matrix ${\bf M}$ for the five-qubit QECC but also the
simplest version of encoder-decoder network corresponding to the derived
matrix. However, many possible matrices ${\bf M}$ suitable for the QECC
remained to be discovered analytically and thus, so many candidates of
encoder-decoder circuit that require only six CNOTs. The simplest network
that is even simpler than the present one and the Braunstein and Smolin
circuit \cite{bs} might not be found from these candidates. However, a more
convincible proof which could be a numerical approach based on the
analytical approach introduced in Ref. \cite{hl} is required in the future
work.

\section{Acknowledgement}

This work is supported partially by the National Science Council, Taiwan
under the grant numbers NSC 94-2212-E-159-002 and NSC 94-2112-M-009-024.

\bigskip

{\LARGE Figure Caption}

\bigskip

Fig. 1. The 1-EPP with notations used in the context. Alice performs $U_{1}$
and ${\frak m}$ and then sends her classical result ($v_{A}$) to Bob. Bob
performs $U_{2}$ and ${\frak m}$, and then combines his own result ($v_{B}$)
and Alice's to control a final operation $U_{3}^{(i)}$.

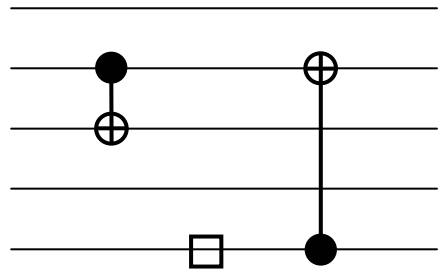
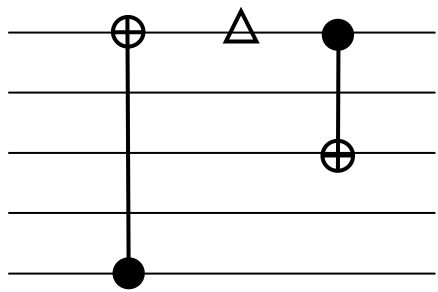
Fig. 2. The three quantum gate arrays performed in the stage of row
operations: (a) for ${\bf M_{1}}\rightarrow {\bf M^{\prime}_{1}}$; (b) for $%
{\bf M^{\prime}_{1}}\rightarrow {\bf M^{\prime\prime}_{1}}$; and (c) for $%
{\bf M^{\prime\prime}_{1}}\rightarrow {\bf 1}$.

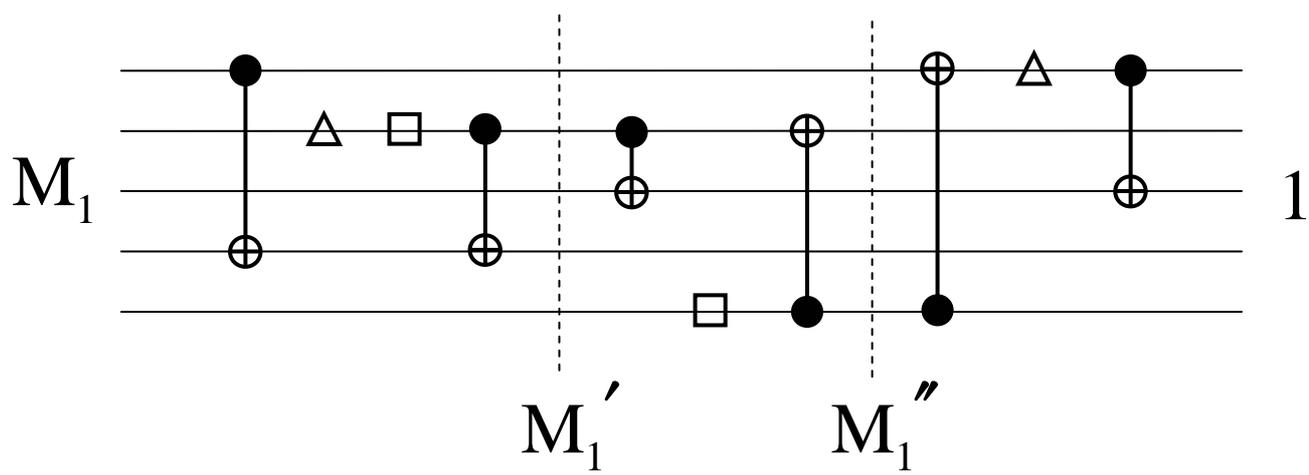
Fig. 3. The gate array for the transformation ${\bf M}_{1}\rightarrow {\bf 1}
$. The basic unitary operations are performed in the order from left to
right, while if they are performed from right to left, then the inverse
transformation ${\bf M_{1}}\rightarrow {\bf 1}$ is accomplished.

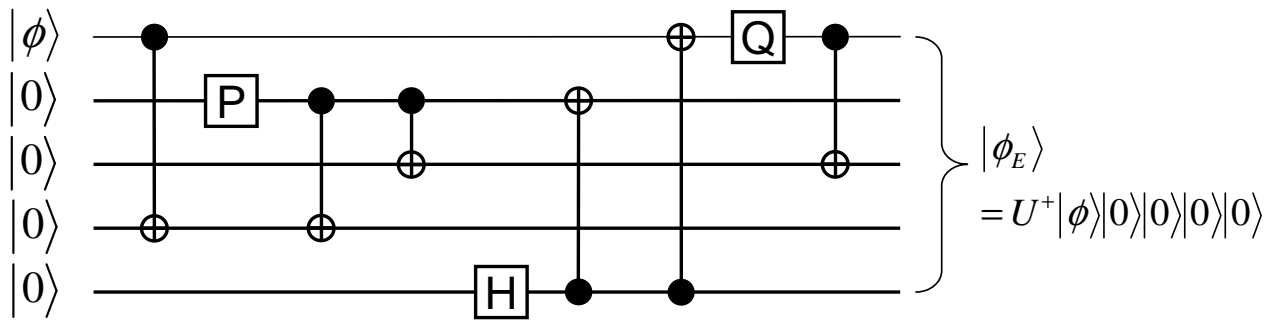
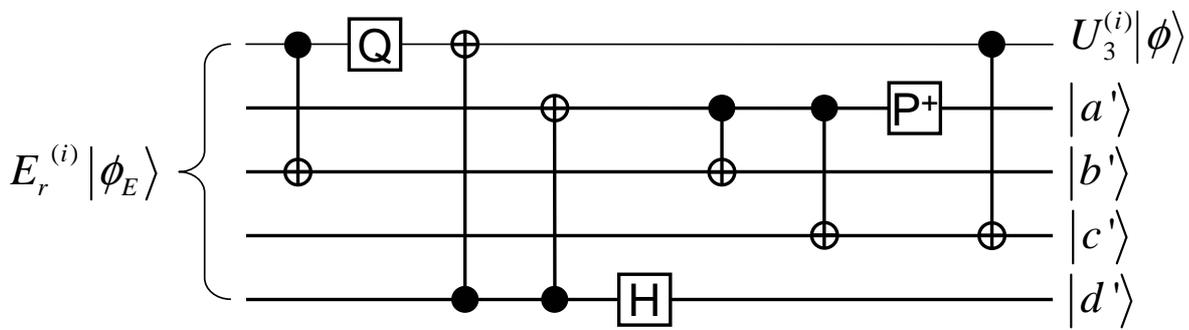
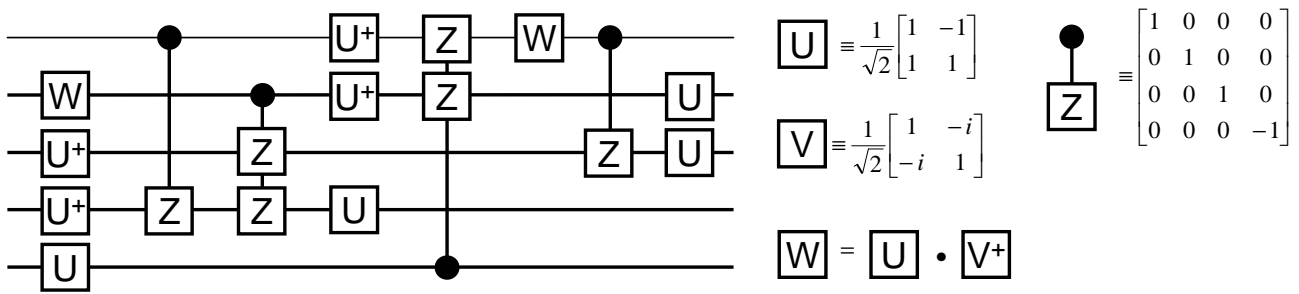
Fig. 4. The perfect five-qubit error correction. (a) The initial tensor
product state is encoded to an entangled state $\left| \phi
_{E}\right\rangle .$ (b) After suffering from the single-qubit error, the
state $E_{r}^{(i)}\left| \phi _{E}\right\rangle $ is then decoded, resulting
in the final tensor product state$(U_{3}^{(i)}\left| \phi\right\rangle
)\left| a^{\prime }b^{\prime }c^{\prime }d^{\prime }\right\rangle .$ Here, $%
{\bf P}=HQ$, ${\bf P}^{+}=QH$. (c) The encoder circuit from Fig. 4(a) is
rewritten in terms of the gate primitives of an ion-trap quantum computer.

\bigskip

{\LARGE Table Caption}

\[
\begin{tabular}[t]{|l|l|l|l|l|}
\hline
$i$ & \ \ \ $e_{r}^{(i)}$, $E_{r}^{(i)}$ & $w^{(i)}$, $W^{(i)}$ & $v^{(i)}$
& $U_{3}^{(i)}$ \\ \hline
0 & 00 00 00 00 00 & 00 00 00 00 00 & 0000 & ${\bf I}$ \\ \hline
1 & 10 00 00 00 00 & 11 00 00 01 01 & 0011 & $\sigma _{y}$ \\ \hline
2 & 01 00 00 00 00 & 01 00 01 01 00 & 0110 & $\sigma _{x}$ \\ \hline
3 & 11 00 00 00 00 & 10 00 01 00 01 & 0101 & $\sigma _{z}$ \\ \hline
4 & 00 10 00 00 00 & 00 01 00 00 01 & 1001 & ${\bf I}$ \\ \hline
5 & 00 01 00 00 00 & 00 11 01 01 00 & 1110 & ${\bf I}$ \\ \hline
6 & 00 11 00 00 00 & 00 10 01 01 01 & 0111 & ${\bf I}$ \\ \hline
7 & 00 00 10 00 00 & 11 01 10 01 01 & 1011 & $\sigma _{y}$ \\ \hline
8 & 00 00 01 00 00 & 00 00 01 00 00 & 0100 & ${\bf I}$ \\ \hline
9 & 00 00 11 00 00 & 11 01 11 01 01 & 1111 & $\sigma _{y}$ \\ \hline
10 & 00 00 00 10 00 & 10 01 00 10 00 & 1000 & $\sigma _{z}$ \\ \hline
11 & 00 00 00 01 00 & 00 00 00 01 00 & 0010 & ${\bf I}$ \\ \hline
12 & 00 00 00 11 00 & 10 01 00 11 00 & 1010 & $\sigma _{z}$ \\ \hline
13 & 00 00 00 00 10 & 00 00 00 00 01 & 0001 & ${\bf I}$ \\ \hline
14 & 00 00 00 00 01 & 01 11 01 00 10 & 1100 & $\sigma _{x}$ \\ \hline
15 & 00 00 00 00 11 & 01 11 01 00 11 & 1101 & $\sigma _{x}$ \\ \hline
\end{tabular}
\bigskip 
\]

Table 1. The correspondence among the error syndrome $e_{r}^{(i)}$ ($%
E_{r}^{(i)}$), the codeword $w^{(i)}$ ($W^{(i)}$), the measurement result $%
v^{(i)},$ and the Pauli operation $U_{3}^{(i)}$ controlled by the
measurement result in the restricted 1-EPP (five-qubit QECC) applying the
encoder-decoder circuit shown in Fig. 3 (Fig. 4)

\[
\begin{tabular}[t]{|l|l|l|l|l|}
\hline
\ \ \ \ \ \ \ \ \ \ \ \ \ {\bf Criteria} & {\bf Circuit 1} & {\bf Circuit 2}
& {\bf Circuit 3} & {\bf Circuit 4} \\ \hline
Total number of operations & 12 & 11 & 10 & 9 \\ \hline
\ \ \ \ \ \ \ Number of CNOT & 7 & 6 & 7 & 6 \\ \hline
\ \ \ \ Number of laser pulse & 35 & * & 26 & 24 \\ \hline
\end{tabular}
\bigskip 
\]

Table 2. Three efficiency criteria and the corresponding costs for four
circuits have been presented. Circuit 1 is given by Bennett {\it et al.}
(Fig. 18 in Ref. {\it \ }\cite{b}) and is unoptimized. The optimized circuit
of Bennett {\it et al}., denoted by Circuit 2, mentioned in Ref. \cite{b}
consists of six two-qubit controlled-NOT gates only. Since the number of
laser pulses depends on the detailed structure of the circuit, it is not
shown here for laking the detailed information. Circuit 3 is the
simplification of the coding circuit of Laflamme {\it et al} proposed by
Braunstein and Smolin (Fig. 1 in Ref. \cite{bs}). One can find that the
original caicuit of Laflamme {\it et al} (Fig. 1 in Ref. \cite{l}) is more
complicated and requires 41 laser pulses. Circuit 4 denotes the simpest
circuit has been found by computer search (Fig. 3 in Ref. \cite{bs}) and by
the systematic method presented in this work.

\end{document}